\documentclass[aps,prl,twocolumn,groupedaddress,showpacs]{revtex4}

\usepackage{graphicx}

\begin{document}


\title{Can Light-nuclei Search Experiments Constrain the Spin-independent Dark Matter Phase Space?}


\author{F. Giuliani}
\email[]{giuliani@phys.unm.edu} \altaffiliation{present address:
Department of Physics and Astronomy, University of New Mexico,
Albuquerque, 87131 NM, USA} \affiliation{Centro de F\'isica Nuclear,
Universidade de Lisboa, 1649-003 Lisboa, Portugal}
\author{TA Girard} \affiliation{Centro de
F\'isica Nuclear, Universidade de Lisboa, 1649-003 Lisboa, Portugal}
\author{T. Morlat}
\altaffiliation{present address: Department of Physics, University
of Montreal, Montreal H3C 3J7, Canada}\affiliation{Centro de
F\'isica Nuclear, Universidade de Lisboa, 1649-003 Lisboa, Portugal}


\date{\today}

\begin{abstract}
At present, restrictions on the spin-independent parameter space of
WIMP dark matter searches have been limited to the results provided
by relatively heavy nuclei experiments, based on the conventional
wisdom that only such experiments can provide significant
spin-independent limits. We examine this wisdom, showing that light
nuclei experiments can in fact provide comparable limits given
comparable exposures, and indicating the potential of light nuclei
detectors to simultaneously and competitively contribute to the
search for both spin-independent and -dependent WIMP dark matter.
\end{abstract}

\pacs{95.35.+d, 29.90.+r }

\maketitle

The direct search for evidence of weakly interacting massive
particle (WIMP) dark matter continues among the forefront activities
of experimental physics. Such searches are traditionally classified
as to whether spin-independent (SI) or spin-dependent (SD),
following the general decomposition of the WIMP-nucleus cross
section into scalar and vector parts. Traditionally, the SI sector
has been the most explored, with the current status of the search
effort defined by a number of detectors which, because of their
target nuclei spins, also provide the defining constraints on the SD
phase space.

Generally, this dual impact is not considered a two-way street: the
prevalent attitude is that exploring the SI sector of WIMP
interactions requires nuclei with a high mass number because of the
coherent enhancement of the scattering cross section, which scales
with the squares of both the mass number and the WIMP-nucleus
reduced mass. This is reflected in the thrust of new search activity
based on detectors with germanium, xenon, cesium, tungsten, and
iodine
\cite{zepmax,papcresstII,libra,edelII,supercdms,hdms,kims,xenon,xmass2,eureka,coupp}.

\begin{figure}[h]
  \includegraphics[width=8 cm]{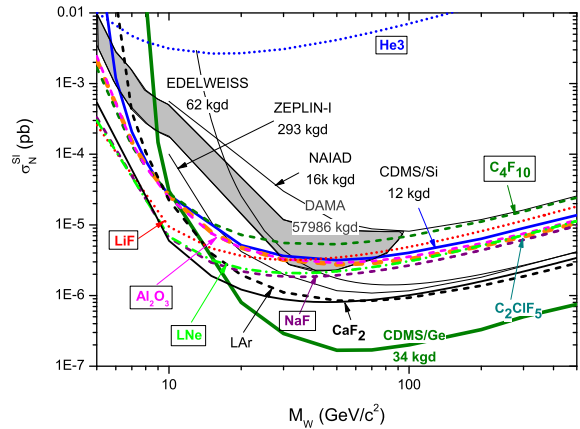}\\
  \caption{projections of SI exclusion contours for various ``light" nuclei
  experiments (broken), with a CDMS/Ge exposure, 8 kev recoil threshold and full
  background discrimination, in comparison with leading experiments (unbroken)
  in this sector.}\label{SI}
\end{figure}

We have examined this conventional wisdom in the case of several
``light" nuclei experiments, which for discussion purposes are
somewhat arbitrarily defined as $A <$ 25.  The main result of this
Letter, which contradicts this ``attitude", is shown in Fig. \ref{SI}. The
reason lies in the loss of coherence in the scattering amplitudes as
a result of the increasing nuclear recoil momentum. This is not to
say that ``heavy" nuclei devices do not provide more restrictive
limits, all else being equal, but only that all is generally not
equal and ``light" nuclei devices cannot be \textit{a priori}
excluded from providing competitive restrictions on the SI parameter
space given comparative exposures and sensitivity.

Fig. \ref{SI} is obtained from straightforward, standard projections of
``light" nuclei results assuming isospin-independence \cite{Lewin}, a
34 kgd exposure (active mass $\times$ live time) equivalent to that
of the current CDMS-II result, and an 8 keV recoil threshold (in
order to focus on only the A-attributable differences). Since
several of the ``light" nuclei experiment techniques also possess
background discrimination capabilities, we further assume for the
purposes of argument that the experiments are able to discriminate
each of the observed events as background: with this assumption, it
can be claimed that no WIMP has been observed, and the 90\% C.L.
upper limit on the WIMP rate is simply $\frac{ln10}{32}$ = 0.068
evt/kgd.

These projections are shown in comparison with several leading SI
experiments \cite{cdms06SI,edel,danai,ZEPLINI,naiad2}, which serve
as benchmarks; the unexcluded region of the controversial positive
DAMA/NaI result is shown as shaded. Although NaI contains sodium, we
do not consider this as ``light" since iodine is also present. This
is similarly true for such other experiments using CaWO$_{4}$
\cite{eureka}, CaF$_{2}$ \cite{CaF}, ZnWO$_{4}$ \cite{eureka} and
CF$_{3}$I \cite{coupp}. Similarly, C$_{2}$ClF$_{5}$ \cite{plb2} is
not considered ``light" because of the chlorine presence, although
the experiment is not generally considered ``heavy".

As seen in Fig. \ref{SI}, several of the indicated ``light" nuclei
experiments would in fact provide restrictions on the parameter
space within an order of magnitude of some of the leading SI
experiments which serve as reference. As also observed, the
unexcluded spin-independent DAMA/NaI region is better probed by the
``light" nuclei than CDMS/Ge.

Although at first sight the ``light" nuclei impact in Fig. \ref{SI} may seem
surprising, it can be understood from the spin-independent WIMP
differential rate on which Fig. \ref{SI} is based \cite{jung,Savage}:

\begin{equation}\label{difrate2}
    \frac{dR}{dE}=\frac{\rho \epsilon(E)}{2M_{W}}\frac{\sigma}{\mu^{2}}F^{2}(q)
    \int_{v_{min}}^{v_{max}}\frac{f(v)}{v}dv  ,
\end{equation}

\noindent where $E$ is the recoil energy of the target nucleus,
$\rho$ is the local WIMP halo mass density, $M_{W}$ is the WIMP
mass, $\sigma$ is the zero-momentum transfer cross section, $\mu$ is
the reduced mass, $F^{2}(q)$ is the nuclear form factor given by
$\frac{\sigma (q)}{\sigma (0)}$ with $q$ the momentum transfer,
$v_{min}$ is the minimum incident WIMP speed required to cause a
recoil of energy $E$, $v_{max}$ is the maximum incident WIMP speed,
$\epsilon$ is the efficiency of the detector, and $f(v)$ is the WIMP
velocity distribution function. The detector-dependent parameters
are $\frac{\sigma}{\mu^{2}}$, $F$, $\epsilon$ and $v_{min}$;
$v_{max}$ is simply the sum of the galactic escape velocity and the
detector velocity with respect to the galaxy.

\begin{figure}[h]
  \includegraphics[width=8 cm]{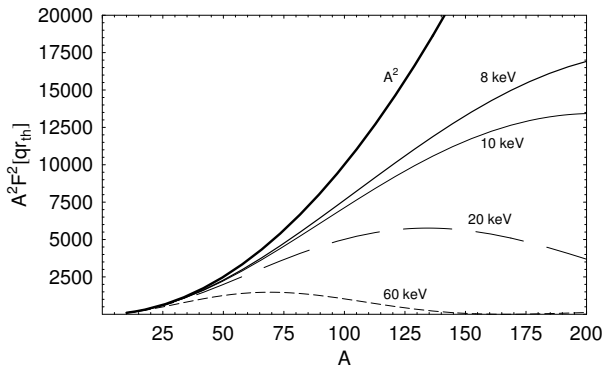}\\
  \caption{Product of $A^{2}F^{2}$ versus $A$,
  calculated for a Helm form factor and various $E$.
  The thick parabola shows the expectation of a simple $A^{2}$ scaling.}
  \label{a2ff}
\end{figure}

The projections of Fig. \ref{SI} employ the standard spherical isothermal
halo, with a local density of 0.3 GeV/c$^{2}$, a halo velocity of
230 km/s, average Earth velocity of 244 km/s, a galactic escape
velocity of 600 km/s, $\epsilon$ = 1 (except in the case of the
superheated liquids where $\epsilon = 1 - \frac{E_{thr}}{E}$, with
$E_{thr}$ the threshold recoil energy), and a Helm form factor
\cite{helmff} ($F(qr_{n}) = 3\frac{j_{l}(qr_{n})}{qr_{n}}
e^{-(qx)^{2}/2}$) with nuclear radius $r_{n} = \sqrt{x^{2} +
\frac{7}{3}\pi^{2}y^{2} - 5z^{2}}$ (x = 1 fm, y = 0.52 fm, z [fm] =
1.23$A^{\frac{1}{3}}- 0.6$) \cite{Lewin}.

As $\frac{\sigma}{\mu^{2}}\propto A^{2}$, the SI exclusion
capability of a nucleus is significantly enhanced by its mass
number, as per conventional wisdom. Moving from $A \sim$ 20 ($F$) to
$A \sim$ 200 ($W$) however only provides a 2 order of magnitude
shift, which is reduced by the higher momentum transfer
corresponding to the high recoil energy bins of large $A$
experiments, unless the analysis is limited to the lowest energies.
Fig. \ref{a2ff} shows the variation in the product $A^{2}F^{2}(q)$ with $A$
for various recoil energies. As evident, for low recoil energies
$A^{2}F^{2}(q)$ increases along the entire range of nuclei, although
above $A \sim$ 40 the growth is weaker than the simple $A^{2}$
scaling. At high recoil energies, the heavy nuclei lose some of
their effectiveness as WIMP detectors. This is particularly evident
in the 20 keV curve, where iodine has a maximum sensitivity. For
iodine recoils in a NaI detector, this curve corresponds to 1.8
keVee, while the 60 keV curve, where iodine has lost most of its
spin-independent sensitivity enhancement in favor of Ge, corresponds
to 5.4 keVee.

\begin{figure}
  \includegraphics[width=8 cm]{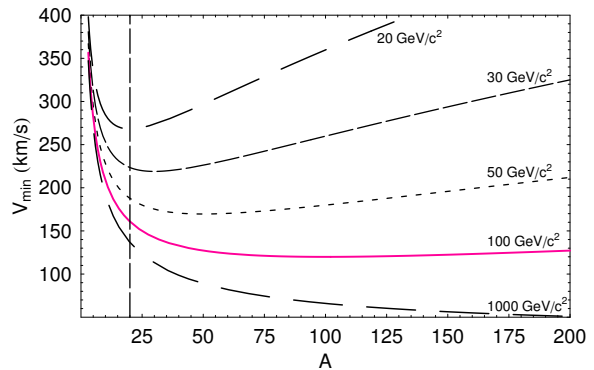}\\
  \caption{Minimum WIMP speed for 8 keV recoil energy and
  various $M_{W}$.}\label{vmin8keV}
\end{figure}

The impact of the integral in Eq. (\ref{difrate2}) on experimental
sensitivities can be discussed in terms of its lower limit:

\begin{equation}\label{vmin}
    v_{min}=\sqrt{\frac{E_{thr}}{2Am_{p}}}(1+\frac{Am_{p}}{M_{W}}) ,
\end{equation}

\noindent \noindent where $m_{p}$ is the proton mass, and the small
difference with the neutron mass is neglected. Clearly, the lower
the $v_{min}$, the larger the inverse velocity integral in Eq. (\ref{difrate2}),
and whenever $v_{min}\geq v_{max}$ the rate vanishes. Hence a WIMP
sensitive experiment should have a low $v_{min}$, otherwise the
detector misses a significant fraction of the recoils within its
sensitive volume, which translates to a loss of constraint even with
full particle discrimination. Low $v_{min}$ can be pursued by
lowering $E_{thr}$, but inspection of Eq. (\ref{vmin}) shows that
the detector composition has a nontrivial effect: $v_{min}$ has an
absolute minimum for $Am_{p}$ = $M_{W}$, so that for $M_{W} \geq
200$ GeV/c$^{2}$ a ``heavy" isotope is an advantage, but for low
$M_{W}$, as visualized in Fig. \ref{vmin8keV} for $E_{thr}$ = 8 keV, this is not
the case.

\begin{figure}[h]
  \includegraphics[width=8 cm]{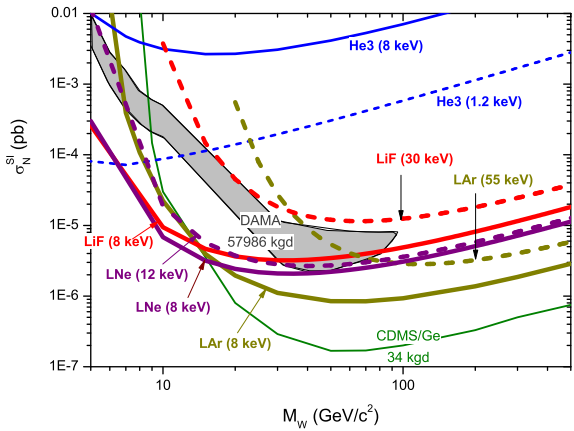}\\
  \caption{comparison of the projected exclusion contours of
  several ``light" nuclei experiments (unbroken) in Fig. \ref{SI} with
  identical projections assuming more realistic recoil threshold
  energies (dashed).}\label{SIthr}
\end{figure}

The strong increase of $v_{min}$ below $A \sim$ 10, visible for all
the displayed WIMP masses, reduces the fraction of the incident WIMP
spectrum detectable by these experiments, unless a strong effort to
lower the recoil threshold below $\approx$ 8 keV (hence to
reduce/discriminate the low energy background) is successfully made.
The question if at low $M_{W}$ a lower inverse velocity integral due
to the choice of ``heavy" nuclei overcomes the benefit from the
$A^{2}F^{2}(q)$ scaling does not have a straightforward,
$f(v)$-independent answer, because this integral depends on the
actual $f(v)$ and $v_{max}$. Assuming the same isothermal halo of
Fig. \ref{SI}, Fig. \ref{SIthr} shows the effect of threshold reduction for the LNe,
LAr, LiF and He3 experiments: clearly, the most stringent contours
for each result from the lowest threshold operation, all else being
equal.

Also observed in Fig. \ref{vmin8keV} is that in a low WIMP mass scenario with
$M_{W} \approx 23$ GeV/c$^{2}$, the DAMA annual modulation signal
can be tested through isotopes in the range $A$ = 15-30. The NaF
bolometers \cite{naf} are probably the most suitable existing
detectors alternative to NaI for this purpose since both nuclei are
Na-similar (the low $M_{W}$ limit of DAMA is dominated by Na).

\begin{table*}
\caption{\label{table}survey of ``light" nuclei experiments,
including projections from new initiatives ($^{\ast}$).}
\begin{ruledtabular}
\begin{tabular}{ccccccc}
  detector & mass & exposure  & $E_{thr}$ & contour min.  & approach & experiment [Ref.]\\
   & (kg) & (kgd) & (keV) & (pb @ GeV/c$^{2}$) & & \\ \hline
  Al$_{2}$O$_{3}$ & 0.262 & 1.51 & $\sim$ 1 & 10$^{-3}$ @ 30 & cryogenic & CRESST-I \cite{cresstI}  \\
          & 0.050 & 0.11 & $\sim$ 2 & 10$^{1}$ @ 50 &  cryogenic & ROSEBUD \cite{rosebudII}  \\
  LiF   & 0.168 & 4.1  & 8 - 61 & 10$^{2}$ @ 30 & cryogenic & Kamioka/LiF \cite{lif}\\
  NaF   & 0.176 & 3.38 & 12 - 37 & 20 @ 20 & cryogenic & Kamioka/NaF \cite{naf} \\
  C$_{4}$F$_{10}$ & 0.019 & 1.98 &  6 & $\sim$ 1 @ 30  & superheated liquid & PICASSO \cite{newpicasso} \\
  $^{\ast}$He3   & 10 & -  & $\sim$ 1  & -  & superfluid & MIMAC \cite{mimac}, ULTIMA \cite{ultima}  \\
  $^{\ast}$LNe   & 5 & -  &  $\sim$ 12 & -  &  noble liquid & CLEAN \cite{clean}  \\
\end{tabular}
\end{ruledtabular}
\end{table*}

Whether or not ``light" nuclei experiments can actually achieve the
larger exposures required to be competitive with the heavier nuclei
devices remains in question, since most have reported only low
active mass, low exposure prototype results, without background
discrimination and with thresholds well above their theoretical
capabilities, as indicated in Table \ref{table}. The searches based on
Al$_{2}$O$_{3}$, LiF, NaF, and He3 are all cryogenic. The CRESST-I
successor, CRESST-II, however pursues CaWO$_{4}$ \cite{eureka}. Both
the LiF and NaF results derived from several bolometers with
thresholds ranging between 8 and 60 keV, the impact of which is
shown in Fig. \ref{SIthr}. Apparently, these resulted from excessive noise
which could not be reduced, and the experiments have been
superceeded by a 310 g CaF$_{2}$ scintillator \cite{CaF}. ROSEBUD,
with plans for a mass increase to 200 g, would require only a 170
kgd exposure to achieve the CDMS contour, which could be
accomplished with a 5 kg active mass in 34 measurement days.
Similarly, the LiF and NaF would require exposure increases to 680
and 450 kgd, respectively, which a few kg active mass could achieve
in reasonable measurement times. The active mass is limited by the
size and cooling power of the refrigeration unit to $\sim$ factor 20
less than the noble gas experiments, and requires some financial
consideration in view of the milliKelvin operating temperatures. All
would profit from background discrimination, using either ionization
or scintillation in addition to heat, with the disadvantage that the
small scintillation signal of nuclear recoils is produced by only
relatively high energy WIMP events.

The C$_{4}$F$_{10}$-based activity is a superheated liquid project,
with a 330 kgd exposure of 4.5 kg active mass currently in progress
which would improve the results to below $10^{-6}$ pb. Such devices
offer an intrinsic insensitivity to the majority of common search
backgrounds, equivalent to an intrinsic rejection factor several
orders of magnitude larger than the bolometer experiments with
particle discrimination; the result is a sensitivity to only high
stopping power interactions, beyond that of nuclear recoils. They
however do not so far offer a background discrimination beyond their
intrinsic insensitivity.

To reach the CDMS contour above $M_{W}$ = 100 GeV/$c^{2}$ would
require of a He3 experiment $>$ 3.4 $\times 10^{4}$ kgd exposure, or
$>$ $10^{3}$ d with the 10 kg device proposed \cite{mimac}. This is
reduced to $\sim$ 1 year if only the $M_{W} <$ 20 GeV/c$^{2}$ region
is in question. Neither of the existing efforts (Table \ref{table}) has yet
produced a result.

The LNe approach is similar to that of xenon (ZEPLIN-II
\cite{zeplin2}, XMASS \cite{xmass2}, XENON \cite{xenon}) and argon
(ArDM \cite{ardm}, DEAP \cite{deap}, WARP \cite{warp}), with a
capacity to identify nuclear recoils. A projected 10 fiducial ton
device is expected to yield a minimum in the limit contour near
10$^{-10}$ pb, but the activity has yet to provide a result. An
advantage of this technique is however that once demonstrated, it
can be rapidly scaled up to significantly larger mass.

In contrast, CDMS is about to start a 5 kg active mass experiment,
which is expected to improve on its current result by an order of
magnitude; this is to be followed by upgrades to 25 kg and
eventually 1 ton \cite{supercdms}. ZEPLIN-II has just reported a 32
kg result, with a contour minimum of $\sim$ 7 $\times$ 10$^{-7}$ pb
at $M_{W}$ = 70 GeV/c$^{2}$; XMASS is preparing a 100 kg active mass
experiment, and XENON is operational with a 15 kg active mass. Of
the LAr experiments, only WARP has produced a first 96 kgd result
with a double phase, 1.83 kg active mass device operating with a
40-55 keV threshold, achieving a minimum of $\sim$ 10$^{-6}$ pb at
90 GeV/c$^{2}$; an upgrade to 186 kg active mass is already
underway. DAMA/LIBRA \cite{libra}, an upgrade of the DAMA/NaI
experiment to 250 kg with improved radiopurity, is running since
2003; R\&D is in progress for a mass upgrade to 1 ton. A second NaI
experiment, ANAIS, reports an exposure of 5.7 kgy with a 10.7 kg
prototype \cite{anais}, and will be eventually upgraded to 100 kg.

In short, whether or not ``light" nuclei experiments will contribute
to the direct search for SI WIMP dark matter depends on whether or
not they are able to close the gap with their more developed, larger
``heavier" colleagues. They however could contribute, given
comparable development and support.

\begin{acknowledgments}
F. Giuliani was supported by grant SFRH/BPD/13995/2003 of the
Portuguese Foundation for Science and Technology (FCT), co-financed
by FEDER; T. Morlat, by grant CFNUL-275-BPD-01/06 of the Nuclear
Physics Center of the University of Lisbon. This study was supported
in part by POCI grant FIS/57834/2004 of FCT, co-financed by FEDER.
\end{acknowledgments}


\end{document}